\documentclass[a4paper]{report}
\usepackage[utf8]{inputenc}
\usepackage[T1]{fontenc}
\usepackage{RJournal}
\usepackage{amsmath,amssymb,array}
\usepackage{booktabs}
\usepackage{hyperref}
\usepackage{verbatim}
\usepackage{cleveref}

\newcommand{\vH}{\boldsymbol{H}}
\newcommand{\vI}{\boldsymbol{I}}
\newcommand{\vJ}{\boldsymbol{J}}

\newcommand{\vP}{\boldsymbol{P}}

\newcommand{\vT}{\boldsymbol{T}}

\newcommand{\vX}{\boldsymbol{X}}
\newcommand{\vY}{\boldsymbol{Y}}
\newcommand{\vZ}{\boldsymbol{Z}}

\newcommand{\Lan}{\mathcal{O}}

\newcommand{\N}{{\mathbb{N}}}

\newcommand{\vmu}{\boldsymbol{\mu}}

\newcommand{\vsigma}{\boldsymbol{\sigma}}
\newcommand{\vSigma}{\boldsymbol{\Sigma}}

\newcommand{\veins}{{\bf 1}}
\newcommand{\vnull}{{\bf 0}}

\DeclareMathOperator{\Var}{Var}

\newcommand{\tr}{\operatorname{tr}}

\newcommand{\bqan}{\begin{eqnarray}}
\newcommand{\eqan}{\end{eqnarray}}

\newcommand{\Hypo}{\mathcal{H}_0}

\newtheorem{theorem}{Theorem}[section]

\newcommand{\rank}{\operatorname{rank}}

\begin{document}

\sectionhead{Contributed research article}
\volume{XX}
\volnumber{YY}
\year{20ZZ}
\month{AAAA}

\begin{article}
\setlength{\parindent}{0pt}
\title{Inference for high dimensional repeated measure designs with the R package hdrm}
\author{by Paavo Sattler and Nils Hichert}
\maketitle
\abstract{Repeated Measure Designs allow comparisons within a group but also between different groups, which is called Split-Plot-Design.  While these designs have a long tradition in agricultural experiments, their utility extends far beyond, including applications in medical research, psychology, and life sciences—fields in which repeated measurements on the same subject are often essential.
In today's age of data collection, where many measurements are taken almost in real time, this often leads to observation vectors whose dimension $d$ exceeds or is at least comparable to the available sample size $N$. Although from a practical perspective, this has advantages in terms of cost-efficiency, ethical considerations, and the investigation of rare diseases, it poses serious methodological challenges for statistical inference. \\
Parametric methods based on the assumption of multivariate normality offer a flexible framework that avoids stringent and often unverifiable conditions on covariance structures or on the asymptotic ratio between $d$ and $N$. 
In such a framework, the freely available R-package \textbf{hdrm} allows the investigation of a multitude of hypotheses regarding the expectation vector for such high-dimensional repeated-measure designs in both single-group and multi-group settings. These include scenarios with homogeneous or heterogeneous covariance matrices. The presented paper describes the tests contained therein, illustrates their usage through extensive examples, and discusses their applicability in practical high-dimensional data settings based on real data. Thereby, homogeneous covariance matrices can ease and improve the estimation of unknown quantities, but are not required. 
To manage computational complexity for the more extensive trace estimators used, especially for large $d$, the package incorporates estimators and subsampling strategies that allow for a substantial reduction in computing time without compromising statistical validity.

}
 \section{Introduction}

Whether for one or several groups, repeated-measure designs are widely used across scientific disciplines wherever the same subjects or experimental units are measured multiple times under different conditions, time points, or locations. This results in observation vectors of dimension $d$, which represent the subjects/objects. There are usually dependencies between these measurements, as they are measured at different times, in different places or under different conditions, but on the same individual. Here, it is often easier to increase the number of measurements $d$ than the sample size $N$.
Consequently, modern repeated-measures data frequently exhibit a high-dimensional character for which conventional statistical methods are not suitable.

 This applies to the investigation of null hypotheses about the expectation vector $\vmu$ for one group as well as for multifactorial designs with $a$ independent groups.
To make meaningful inference possible, further assumptions must therefore be made. A common restriction of existing methods is requirements on the relation between $d$ and $N$, such as $d^3/N\to 0$ as in \cite{huber1973}, $d/N\to c\in(0,1)$ as in \cite{bai1996} or $d/N\to c\in(0,\infty)$ as in \cite{harrar2016}. In \cite{happ2016}, this was further enhanced by imposing multiple conditions on the relation between $d$ and traces that contain the unknown covariances. The resulting tests are part of the R Package \textbf{HRM} \citep{HRM}, which was described in \cite{hRM_RJournal}.\\  However, since dimension and sample sizes are known in any given application, while their asymptotic behaviour is not, these assumptions are often problematic to validate in practice.
Furthermore, for many procedures, special conditions on the underlying covariance matrices, such as some kind of sparsity, are usual, as in \cite{cai2013}. For the case of $a=2$ and a special class of hypotheses, the work of \cite{chen2010} is a notable exception, which does without assumptions on the relation between $d$ and $N$ or regarding the covariance matrix. If a parameter $\tau_{CQ}$, which is a ratio of different traces containing the unknown covariance matrices, converges to zero, this leads to an asymptotic correct test, which is part of multiple existing R packages, like \textbf{PEtests} (\cite{PEtests}) or \textbf{highD2pop} (\cite{highD2pop}).\\\\
For one group and multivariate observation vectors, \cite{pauly2015} developed an asymptotic test that performs well even in small samples and is applicable for $\tau_{CQ}\to c\in\{0,1\}$. 
This approach was generalised by \cite{sattler2018} to accommodate an arbitrary number of groups and multiple asymptotic frameworks, including those with an increasing number of groups, as previously considered in \cite{bathke2008}, \cite{bathke2008a}, and \cite{zhan2014}. This is especially useful if there are multiple factors by which subjects can be separated.
In \cite{sattler2021}, these frameworks and results are further extended in the case of homogeneous covariance matrices.

The R-package \textbf{hdrm} \cite{hdrm2} for high-dimensional repeated measures contains the test procedures from all these three papers to provide consistent tests for one-group setting as well as for multiple groups in heterogeneous and homogeneous covariance settings.The package is currently available on GitHub at
\url{https://github.com/Schnieboli/hdrm} providing easy and uncomplicated access for users; a CRAN submission is planned. It contains the two main functions \textit{hdrm\_single} and \textit{hdrm\_grouped} for the one-group, respectively, multiple-group case, where the latter contains optional arguments to specify equality of covariance matrices.\\\\ 
In the following section, we will introduce the statistical multivariate model and the null hypotheses being investigated, along with the underlying assumptions. We will then present the inference procedures that were implemented, followed by applications from the R package \textbf{hdrm}, demonstrated with several illustrative examples. Finally, we will conclude the paper with a summary and an outlook on potential future developments.

 \section{Statistical Model and Hypotheses}
We consider a split-plot design comprising $a$ independent groups of $d$-dimensional random vectors given through
\bqan\label{eq:model}
{\vX}_{i,j}= ({X}_{i,j,1},\dots,{X}_{i,j,d})^\top \stackrel{ind}{\sim}\mathcal{N}_d\left(\vmu_i,\vSigma_i\right)\hspace{0,2cm}j=1,\dots,n_i,\hspace{0,2cm} i=1,\dots,a
\eqan
where each $\vX_{i,j} \in \mathbb{R}^d$ denotes the observation vector from subject $j$ in group $i$, with mean vector $\vmu_i = (\mu_{i,t})_{t=1}^d \in \mathbb{R}^d$ and positive-definite covariance matrix $\vSigma_i > 0$.

Let $\vmu = (\vmu_1^\top, \dots, \vmu_a^\top)^\top \in \mathbb{R}^{ad}$ denote the overall mean vector. Hypotheses concerning linear combinations of group means can be expressed in the general form
\[
\mathcal{H}(\vH):\ \vH \vmu = \mathbf{0}_{ad},
\]
where the hypothesis matrix $\vH$ is assumed to have the Kronecker product structure $\vH = \vH_W \otimes \vH_S$, with $\vH_W \in \mathbb{R}^{a \times a}$ and $\vH_S \in \mathbb{R}^{d \times d}$ representing the group-level (whole-plot) and time-level (subplot) structures, respectively.

Since the same hypothesis may be expressed through different matrices $\vH$, it is natural to consider the associated unique projection matrix
\[
\vT = \vH^\top (\vH \vH^\top)^+ \vH,
\]
which uses the Moore-Penrose inverse and satisfies $\vH \vmu = \mathbf{0} \iff \vT \vmu = \mathbf{0}$. This matrix is invariant under different matrix representations of the hypothesis (see \cite{brunner2019}). Furthermore, due to the Kronecker structure, the projection matrix $\vT$ also decomposes as $\vT = \vT_W \otimes \vT_S$, where\[
\vT_W = \vH_W^\top (\vH_W \vH_W^\top)^+ \vH_W \quad \text{and} \quad \vT_S = \vH_S^\top (\vH_S \vH_S^\top)^+ \vH_S.
\]

Several common hypotheses in repeated measures and split-plot designs can be formulated using this structure, for example:

\begin{itemize}

\item[(a)] No whole plot(group) effect:\\ $\Hypo^{Whole}: \left(\vP_a\otimes \frac{1}{d}\vJ_d\right)\vmu=\vnull$,\vspace{0.1cm}
\item[(b)] No subplot(time) effect:\\ $\Hypo^{Subplot}: \left(\frac 1 a\vJ_a\otimes \vP_d\right)\vmu=\vnull$,\vspace{0.1cm}
\item[(c)] No interaction effect between whole plot and subplot:\\
$\Hypo^{Interaction}: \left(\vP_a\otimes \vP_d\right)\vmu=\vnull$,\vspace{0.1cm}
\item[(d)] Identical group expectations:\\
$\Hypo^{Identical}: \left(\vP_a\otimes \vI_d\right)\vmu=\vnull$,\vspace{0.1cm}
\item[(e)] Flat time profile:\\
$\Hypo^{Flat}: \left(\vI_a\otimes \vP_d\right)\vmu=\vnull$.
\end{itemize}
Here, $\vP_d = \vI_d - \frac{1}{d} \vJ_d$ denotes the centering matrix (projection onto the space orthogonal to the constant vector), $\vI_d$ the identity and $\vJ_d = \veins_d \veins_d^\top$ is the all-ones matrix.\\

For all these hypotheses $\Hypo(\vT)$, we derive asymptotic tests in settings where not only the sample sizes but also the dimension and the number of groups can increase.
To be concrete, for a heterogeneous covariance setting, we consider three different asymptotic frameworks, which are:

\renewcommand{\theequation}{\Roman{equation}}
\setcounter{equation}{0}
\begin{eqnarray}
a,\min(n_1,...,n_a)\to \infty,\label{asframe1}\\
d,\min(n_1,...,n_a)\to \infty,\label{asframe2}\\
a,d,\min(n_1,...,n_a) \to \infty\label{asframe3}.
\end{eqnarray}
Moreover, here one assumption has to be made about the relation between the group sample sizes $n_i$ and the total sample size, $N=\sum_{i=1}^a n_i$.
Therefore, for the above settings, we assume that for all $i=1,...a$ it holds $n_i/N\to\kappa_{i}\in (0,1]$, which ensures that no group becomes asymptotically negligible.
In case of homogeneous covariance matrices $\vSigma_1=...=\vSigma_a=\vSigma$, an even wider range of asymptotic frameworks can be considered, through
\begin{eqnarray}
a\to \infty,\label{asframe4}\\
a,d\to \infty,\label{asframe5}\\
a,\max(n_1,...,n_a)\to \infty,\label{asframe6}\\
d,\max(n_1,...,n_a)\to \infty,\label{asframe7}\\
a,d,\max(n_1,...,n_a) \to \infty\label{asframe8}.
\end{eqnarray}
\renewcommand{\theequation}{\arabic{equation}}
\setcounter{equation}{2}
Notably, these frameworks do not impose any fixed relation between $a$, $d$, and $n_i$, except that the total sample size satisfies $N \to \infty$ in all cases.

 \section{Inference}

To investigate these null hypotheses, we consider group means $\overline {\vX}_{i} = n_i^{-1} \sum_{j=1}^{n_i} \vX_{i,j}, i=1,\dots,a,$ and the pooled mean vector 
${\overline{\vX}^\top=(\overline\vX_1^\top,\dots\overline\vX_a^\top)}$. Based on this vector, we construct the quadratic form 
\bqan\label{eq: stat}
Q_N=N\cdot\overline {\vX}^\top \vT\overline {\vX},
\eqan
commonly referred to as the \emph{ANOVA-type statistic} (ATS), originally proposed in \cite{brunner1997}. However, since, depending on the underlying unknown covariance matrices, $Q_N$ and may diverge to infinity with increasing dimension, we instead use the standardized and centred version $\widetilde W_N  =\{Q_N-\E_{\Hypo}(Q_N)\}/\Var_{\Hypo}\left(Q_N\right)^{1/2}$ which enables asymptotically valid inference under various high-dimensional settings. To compute the expectation and variance of $Q_N$ under the null hypothesis, we define the block-diagonal matrix
$\Var(\sqrt{N}\ \overline\vX)=\vSigma_N=\bigoplus_{i=1}^a \frac N{n_i} \vSigma_i$, while $\bigoplus$ denotes the direct sum, expectation value and variance of the quadratic form can be calculated as follows:


\begin{align}\label{eq: muQ1}
\E\nolimits_{\Hypo}\left(Q_N\right)&=\tr\left(\vT \vSigma_N \right)=\sum_{i=1}^a \frac{N}{n_i}(\vT_W)_{ii}\tr\left(\vT_S\vSigma_i\right),
\end{align}

\begin{align}\nonumber
\Var_{\Hypo}\left(Q_N\right)&
  = \  2\tr\left(\left(\vT \vSigma_N \right)^2\right)\\
&= \ 4\sum\limits_{i,r=1, r<i}^a  \frac{N^2}{n_in_r}{(\vT_W)_{ir}}^2\tr\left(\vT_S\vSigma_i\vT_S\vSigma_r\right)+2\sum\limits_{i=1}^a  \frac{N^2}{n_i^2}{(\vT_W)_{ii}}^2\tr\left(\left(\vT_S\vSigma_i\right)^2\right).
\label{QNVar1}\end{align}
The matrix product $\vT \vSigma_N\vT$ plays a central role, not only for calculating these moments but also for determining the asymptotic distribution of $\widetilde{W}_N$.
Specifically, the limiting distribution essentially depends on $\lambda_1,...,\lambda_{ad}$, which are the eigenvalues of $\vT\Sigma_N\vT$ in decreasing order, as formalized in the following result from \cite{sattler2018} :

\begin{theorem}\label{Theorem3}

Let $\beta_s={\lambda_s}\Big/{\sqrt{\sum_{\ell=1}^{ad}\lambda_\ell^2}}$ for $s=1,\dots,ad$. 
Then $\widetilde W_N$ has, under $\Hypo(\vT)$, and one of the 
frameworks \eqref{asframe1}-\eqref{asframe8} asymptotically
\begin{itemize} 
\item[a)]a standard normal distribution {if and only if}
\[\beta_1 = \max_{s\leq ad} \beta_{s} \to 0 \hspace{0.5cm} \text{as}\hspace{0.2cm} N \to \infty,\]
\item[b)]a standardized $\left(\chi_1^2-1\right)/\sqrt{2}$ distribution {if and only if}
\[\beta_1\to 1 \hspace{0.5cm} \text{as}\hspace{0.2cm} N \to \infty,\]
\end{itemize}
\end{theorem}
Note that convergence to the limiting distribution in frameworks \eqref{asframe4}--\eqref{asframe8} implicitly assumes homogeneous covariance matrices.\\

A major challenge for creating an asymptotically correct level $\alpha$ test is that the limiting distribution changes with the underlying eigenvalue structure, i.e., depending on whether $\beta_1 \to 0$ or $\beta_1 \to 1$. Nonetheless, both situations can be addressed uniformly using the \emph{Pearson approximation}, introduced in \cite{pauly2015}. This approximation constructs a random vairbale $K_{f_P}$ matching the first three moments of $\widetilde{W}_N$, where
\begin{equation}\label{Approximation}
K_{f_P} := \frac{\chi^2_{f_P} - f_P}{\sqrt{2f_P}}, \quad \text{with } \quad f_P = \frac{\tr\left((\vT \vSigma_N)^2\right)^3}{\tr\left((\vT \vSigma_N)^3\right)^2}.
\end{equation}

\begin{theorem}\label{Theorem5}
Under the assumptions of Theorem~\ref{Theorem3} and any of the asymptotic frameworks \eqref{asframe1}--\eqref{asframe8}, the Pearson-approximating statistic $K_{f_P}$ converges in distribution under $\mathcal{H}(\vT)$ as follows:
\begin{itemize} 
\item[a)]a standard normal distribution  if $\beta_1\to 0$ as $N \to \infty$,
\item[b)]a standardized $\left(\chi_1^2-1\right)/\sqrt{2}$ distribution if $\beta_1\to 1$ as $N \to \infty$.
\end{itemize}
\end{theorem}
The parameter $\tau_P:= 1/f_P$ can be interpreted as a convergence index and makes a connection between $\beta_1$ and $f_P$.
 It holds that $\tau_P \to 0$ if and only if $\beta_1 \to 0$, and $\tau_P \to 1$ if and only if $\beta_1 \to 1$. Thus, quantiles based on the Pearson approximation $K_{f_P}$ provide a unified opportunity for critical value determination, regardless of whether $\beta_1\to 0$ or $\beta_1\to 1$.

 In practice, the implementation of this approach requires consistent estimators for the trace terms involved, which depend on the underlying setting. These estimation procedures are described in the next section and are also integrated into the R package \textbf{hdrm}.

 \subsection{One Group}\label{OG} 
In the special case of a single group ($a = 1$), the pooled covariance matrix simplifies to $\vSigma_N = \vSigma_1$. In this setting, unbiased and ratio-consistent estimators for the trace terms involving $\vSigma_1$ can be constructed using symmetrized $U$-statistics based on quadratic forms. Following the approach developed in \cite{pauly2015}, we define the following estimators:

 \[\begin{array}{l}
A_1=\frac{1}{N}\sum\limits_{\ell=1}^N \vX_{1,\ell}^\top \vT\vX_{1,\ell}
\\[4.0ex]
A_2=\frac{1}{2\cdot N(N-1)} \sum\limits_{\begin{footnotesize}\substack{\ell_1, \ell_2=1\\ \ell_1>\ell_2}\end{footnotesize}}^{N} \vX_{1,\ell_1}^\top \vT\vX_{1,\ell_2}
\\[4.0ex]
A_3=\binom{N}{3}^{-1} \sum\limits_{\ell_1=1}^{N-2}\sum\limits_{\ell_2=\ell_1+1}^{N-1}\sum\limits_{\ell_3=\ell_2+1}^N \vX_{1,\ell_1}^\top \vT\vX_{1,\ell_2}\cdot
\vX_{1,\ell_2}^\top \vT\vX_{1,\ell_3}\cdot \vX_{1,\ell_3}^\top \vT\vX_{1,\ell_1}
\end{array}\]

These estimators all use the null hypothesis, which allows them to be unbiased and ratio-consistent for the trace expressions $\tr(\vT\vSigma_1)$, $\tr((\vT\vSigma_1)^2)$ and $\tr((\vT\vSigma_1)^3)$.
As a result, under $\Hypo\vT\vmu=\vnull$, the degrees of freedom parameter $f_P$ from the Pearson approximation can be consistently estimated by
 $\widehat f_P^A:=A_2^3/A_3^2$.\\
So, the resulting test statistic incorporating these estimators is given by

\[W_N^A=\frac{Q_N-A_1}{\sqrt{2 A_2}}.\]

Under the null hypothesis $ \Hypo(\vT) $ and $ \beta_1 \to  b_1\in \{0, 1\}$, we have the following results for the asymptotic framework specified in  equation \eqref{asframe4}:

\[\begin{array}{ll}
\widetilde W_N- W_N^A&=\Lan_P(1)\\
\widetilde W_N- K_{\widehat f_P^A}&=\Lan_P(1)\end{array}\]
Consequently, with the above conditions, an asymptotic correct level $\alpha$ test  for the null hypothesis $\vH(\vT)$ is given through
$\varphi_N^A = \mathbf{1}\{W_N^A > K_{\hat{f}_P^A; 1-\alpha}\}$, with $K_{\hat{f}_P; 1-\alpha}$ denoting the $(1-\alpha)$ quantile of the approximating distribution $K_{\widehat f_P^A}$.
 
 \subsection{Multi-factorial}

 The decomposition of the expectation and variance of the test statistic $Q_N$ given in \Cref{eq: muQ1} and \Cref{QNVar1} allows us to estimate the unknown traces through multiple estimators for the single elements. This approach is not only mathematically more robust — since it leverages the null hypothesis $\vT\vmu = \vnull$ — but also preferable from a practical standpoint. 
 To this end, we utilize within-group differences defined through $\vY_{i,\ell_1,\ell_2}:=(\vX_{i,\ell_1}-\vX_{i,\ell_2})$ which are distributed as $\vY_{i,\ell_1,\ell_2}\sim\mathcal{N}_d(\vnull_d,2\vSigma_i)$ for $\ell_1\neq \ell_2$. Based on these vectors, the single-group estimators from the previous subsection can be adapted, resulting in 
\[\begin{array}{l}
B_{i,1}=\frac 1 {2\cdot \binom {n_i} {2}} \sum\limits_{\begin{footnotesize}\substack{\ell_1,\ell_2=1\\ \ell_1>\ell_2}\end{footnotesize}}^{n_i} \vY_{i,\ell_1,\ell_2}^\top \vT_S\vY_{i,\ell_1,\ell_2},
\\[4.0ex]
B_{2}=\sum\limits_{i=1}^a \frac{N}{n_i } (\vT_W)_{ii} B_{i,1},
\\[4.0ex]
B_{i,r,3}=\frac  1 {4\cdot\binom {n_i} {2}\binom {n_r} {2}} \sum\limits_{\begin{footnotesize}\substack{\ell_1,\ell_2=1\\ \ell_1>\ell_2}\end{footnotesize}}^{n_i}\sum\limits_{\begin{footnotesize}\substack{k_1,k_2=1\\ k_1>k_2}\end{footnotesize}}^{n_r}\left[\vY_{i,\ell_1,\ell_2}^\top \vT_S\vY_{r,k_1,k_2}\right]^2,
\\[4.0ex]
B_{i,4}=\frac 1 {4\cdot 6\binom {n_i} {4}} \sum\limits_{\begin{footnotesize}\substack{\ell_1,\ell_2=1\\ \ell_1>\ell_2}\end{footnotesize}}^{n_i}
\sum\limits_{\begin{footnotesize}\substack{k_2=1\\k_2\neq \ell_1\neq \ell_2 }\end{footnotesize}}^{n_i}\sum\limits_{\begin{footnotesize}\substack{k_1=1\\ \ell_2\neq \ell_1\neq k_1>k_2}\end{footnotesize}}^{n_i}
\left[\vY_{i,\ell_1,\ell_2}^\top \vT_S\vY_{i,k_1,k_2}\right]^2,
\\[4.3ex]
B_5=\sum\limits_{i=1}^a \left(\frac{N}{n_i }\right)^2 {(\vT_W)_{ii}}^2 B_{i,4}+2\sum\limits_{i=1}^a\sum\limits_{\begin{footnotesize}\substack{r,i=1\\ r<i}\end{footnotesize}}^a \frac{N^2}{n_in_r } {(\vT_W)_{ir}}^2 B_{i,r,3}.
\end{array}\]

Here, of course, more traces have to be estimated, and linear combinations of them are necessary to estimate the variance of $Q_N$. Moreover, building the differences results in many more possible index combinations and, therefore, a higher number of summations, increasing the computational effort accordingly.\\

To estimate the third trace $\tr((\vT\vSigma_N)^3)$, a decomposition in smaller parts is unfortunately not feasible, such that another approach has to be used. To this aim, we adapt $A_3$ for a more complicated random vector. To this end, we introduce joint normal random vectors
\[\vZ_{(\ell_1,\ell_2,\dots,\ell_{2a})}:=\left(\sqrt{\frac{N}{n_1}}\vY_{1,\ell_1,\ell_2}^\top\textbf{,}\dots\textbf{,}\sqrt{\frac{N}{n_a}}\vY_{a,\ell_{2a-1},\ell_{2a}}^\top\right)^\top \] are introduced, with $1\leq \ell_{2i-1}\neq \ell_{2i}\leq n_i$ for all $i=1\dots,a$. These random vectors of course requires $n_i\geq 6$ and fullfill $\vZ_{(\ell_1,\ell_2,\dots,\ell_{2a})}\sim \mathcal{N}_{ad}(\vnull,2\vSigma_N)$.

With $\ell_1\neq \ell_2\neq ... \neq \ell_6$ means that all indices are different, this leads to the definition of
\bqan\label{eq:C5}
B_{6}=\hspace*{-0.2cm}\sum\limits_{\begin{footnotesize}\substack{\ell_{1,1},\dots, \ell_{6,1}=1\\
\ell_{1,1}\neq\dots\neq \ell_{6,1}}\end{footnotesize}
}^{n_1}\dots\hspace*{-0.2cm}\sum\limits_{\begin{footnotesize}\substack{\ell_{1,a},\dots, \ell_{6,a}=1\\
\ell_{1,a}\neq\dots\neq \ell_{6,a}}\end{footnotesize}
}^{n_a}
\frac{\Lambda_{1}(\ell_{1,1},\dots,\ell_{6,a})\cdot \Lambda_{2}(\ell_{1,1},\dots,\ell_{6,a})\cdot \Lambda_{3}(\ell_{1,1},\dots,\ell_{6,a})}{8\cdot \prod\limits_{i=1}^a\frac {  n_i! }{\left(n_i-6\right)!}},
\eqan
with the auxilary terms
\[\Lambda_{1}(\ell_{1,1},\dots,\ell_{6,a})={\vZ_{(\ell_{1,1},\ell_{2,1},\dots,\ell_{1,a},\ell_{2,a})}}^\top \vT \vZ_{(\ell_{3,1},\ell_{4,1},\dots,\ell_{3,a},\ell_{4,a})},\]
\[\Lambda_{2}(\ell_{1,1},\dots,\ell_{6,a})={\vZ_{(\ell_{3,1},\ell_{4,1},\dots,\ell_{3,a},\ell_{4,a})}}^\top \vT \vZ_{(\ell_{5,1},\ell_{6,1},\dots,\ell_{5,a},\ell_{6,a})}, \]
\[\Lambda_{3}(\ell_{1,1},\dots,\ell_{6,a})={\vZ_{(\ell_{5,1},\ell_{6,1},\dots,\ell_{5,a},\ell_{6,a})}}^\top \vT \vZ_{(\ell_{1,1},\ell_{2,1},\dots,\ell_{1,a},\ell_{2,a})}.\]\\
Although this estimator has all desired properties, from a practical perspective, it is less applicable since its computation involves $\prod_{i=1}^a\tfrac {  n_i! }{\left(n_i-6\right)!}$ summations. This results in a large computation time, such that this estimator has only limited usability in many settings. To overcome this issue, we introduce a subsampling version of this estimator, which allows us to control the number of calculated summations.
Thereto, instead of considering all possible index combinations, we consider only $B\in\N$ randomly drawn index combinations.
                                                                                            
So for each $i=1,\dots,a$ and $b=1,\dots,B$ subsamples $\{\sigma_{1i}(b),\dots,\sigma_{6i}(b)\}$ of length $6$ were drawn independly from $\{1,\dots,n_i\}$. Then using a joint random vector $\vsigma(b) = (\sigma_{11}(b),\dots,\sigma_{6a}(b))$, the subsampling version of $B_6$ is given through

\[{B_6^\star} = {B_6^\star}\left(B\right)=\frac 1 {8 \cdot B}\sum\limits_{b=1}^B 
 \Lambda_1(\vsigma(b)) \cdot \Lambda_2(\vsigma(b)) \cdot \Lambda_3(\vsigma(b)). 
\]
Now, if $B\to \infty$ when $N\to \infty$, this is an unbiased and ratio-consistent estimator for $\tr((\vT\vSigma_N)^3)$, which can be assured, for example, through polynomial functions of $N$. This results in the final test statistic \[W_N^B=\frac{Q_N-B_2}{\sqrt{2 B_5}}\] and ratio-consistent estimator $\widehat f_P^B:=B_5^3/(B_6^\star)^2$ for the degree of freedom.

Then under $\Hypo(\vT)$, for $\beta_1\to b_1\in\{0,1\}$ and one of the asymptotic frameworks \eqref{asframe1}- \eqref{asframe3} we obtain

\[\begin{array}{ll}
\widetilde W_N- W_N^B&=\Lan_P(1)\\
 W_N^B- K_{\widehat f_P^B}&=\Lan_P(1),\end{array}\]
 yielding an asymptotic correct level $\alpha$ test for $\Hypo(\vT)$ through $\varphi_N^B = \mathbf{1}\{W_N^B > K_{\hat{f}_P^B; 1-\alpha}\}$.\\\\

In practice, large group sizes may also make the computation of $B_{i,1}, B_{i,r,3}$ and $B_{i,4}$ infeasible. In these cases, subsampling versions thereof can also be useful to reduce computational burden while preserving consistency. Thereto, analogous $B$ random subsamples are drawn, and the quadratic forms are only calculated for them.

\subsection{Homogenous Covariance Matrices}\label{HCM}
In the case we have a homogenous covariance matrix setting with $\vSigma_1=...=\vSigma_a=\vSigma$, this can be used to simplify the expectation and variance of the quadratic form $Q_N$ considerably
\[\begin{array}{ll}\E_{\Hypo}(Q_N)&=\tr(\vT_S\vSigma)\cdot \sum\limits_{i=1}^a \frac{N}{n_i} (\vT_W)_{ii}\\[1.4ex]
\Var_{\Hypo}(Q_N)&=2\cdot \tr((\vT_S\vSigma)^2)\cdot \sum\limits_{i=1}^a\sum\limits_{r=1}^a \frac{N^2}{n_i n_r} (\vT_W)_{ir}^2.\end{array}\]
As a result, only the trace terms $\tr(\vT_S\vSigma)$ and $\tr((\vT_S\vSigma)^2)$ need to be estimated, whereto data from all groups can be pooled to construct unified estimators.

In analogy to the one-group case from \Cref{OG}
estimators based on differences $Y_{i,\ell_1,\ell_2}$ are calculated and afterwards are averaged over all groups. This leads to estimators:
\[\begin{array}{l}
C_1=\frac 1 { \sum_{i=1}^a (n_i-1)n_i}\sum\limits_{i=1}^a  \sum\limits_{\ell_1< \ell_2=1}^{n_i} \vY_{i,\ell_1,\ell_2}^\top\vT_S \vY_{i,\ell_1,\ell_2}\\[4ex]

C_2=\sum\limits_{i=1}^a \sum\limits_{\begin{footnotesize}\substack{\ell_1,\ell_2=1\\ \ell_1>\ell_2}\end{footnotesize}}^{n_i}
\sum\limits_{\begin{footnotesize}\substack{k_2=1\\k_2\neq \ell_1\neq \ell_2 }\end{footnotesize}}^{n_i}\sum\limits_{\begin{footnotesize}\substack{k_1=1\\ \ell_2\neq \ell_1\neq k_1>k_2}\end{footnotesize}}^{n_i}
\frac{\left[\vY_{i,\ell_1,\ell_2}^\top \vT_S\vY_{i,k_1,k_2}\right]^2} {4\cdot 6 \sum_{i=1}^a \binom{n_i} {4}}
\\[4ex]

C_{3}:=\frac{1}{6! \cdot \sum_{j=1}^a \binom{n_j}{6}}\sum\limits_{i=1}^a \left(\frac{1}{8}\sum\limits_{\ell_1\neq ...\neq \ell_6=1}^{n_i} \vY_{i,\ell_1,\ell_2}^\top \vT_S\vY_{i,\ell_3,\ell_4}\cdot \vY_{i,\ell_3,\ell_4}^\top\vT_S \vY_{i,\ell_5,\ell_6}\cdot \vY_{i,\ell_5,\ell_6}^\top\vT_S \vY_{i,\ell_1,\ell_2}\right).\end{array}\] 
Again, the number of summations for the corresponding estimator $C_3$ is very high, making it computationally intensive. A subsampling version of this estimator again addresses this issue and  is given through
\[C_{3}^\star:=\frac{1}{a\cdot B}\sum\limits_{i=1}^a \sum\limits_{b=1}^B \vY_{i,\sigma_{i1}(b),\sigma_{i2}(b)}^\top \vT_S\vY_{i,\sigma_{i3}(b),\sigma_{i4}(b)}\cdot \vY_{i,\sigma_{i3}(b),\sigma_{i4}(b)}^\top\vT_S \vY_{i,\sigma_{i5}(b),\sigma_{i6}(b)}\cdot \vY_{i,\sigma_{i5}(b),\sigma_{i5}(b)}^\top\vT_S \vY_{i,\sigma_{i1}(b),\sigma_{i2}(b)}\]
which is unbiased and ratio consistent for $\tr((\vT_S\vSigma)^3)$, if $B\to \infty$ when $N\to \infty$.  This leads to a ratio-consistent estimator $\widehat f_P^C:=C_2^3/(C_3^\star)^2$ for the degree of freedom and the test statistic \[W_N^C=\frac{Q_N-C_1}{\sqrt{2 C_2}}.\]

Now, under the null hypothesis $ \Hypo(\vT) $, for $\beta_1\to b_1\in\{0, 1\}$, and all asymptotic frameworks \eqref{asframe4}-\eqref{asframe8} it holds

\[\begin{array}{ll}
\widetilde W_N- W_N^C&=\Lan_P(1)\\
 W_N^C- K_{\widehat f_P^C}&=\Lan_P(1)\end{array}\]
resulting in an asymptotic correct level $\alpha$ test for $\Hypo(\vT)$ through $\varphi_N^C = \mathbf{1}\{W_N^C > K_{\hat{f}_P^C; 1-\alpha}\}$.

\subsection{Alternative Hypothesis matrices for computation effort reduction}
The computational effort and, consequently, the computation time are main criteria for the applicability of estimators in high-dimensional settings, as calculations become increasingly demanding with increasing dimension~$d$. Therefore, in addition to employing subsampling estimators, further strategies are implemented to reduce the computational burden of the proposed test procedure. A key aspect in this context is the dimension of the hypothesis matrix, which is used within each quadratic form. As illustrated in the examples, such matrices often have a rank that is substantially smaller than~$ad$; for instance, it holds that  $\rank(\vP_d)=d-1$. In \cite{sattler2025}, building on the results of \cite{sattler2023}, the authors show the existence of an alternative hypothesis matrix with a minimal number of rows that allows the same hypothesis to be formulated without changing the value of the test statistic.
Their application, therefore, can lead to a substantial reduction of time for all estimators. Since this so-called companion matrix exists for each possible hypothesis matrix, it is strongly recommended to apply this reduction whenever the rank is clearly smaller than the matrix dimension.

\section{Software and examples}
In this section, we examine data sets to test several hypotheses, demonstrating the application of both test functions implemented in the \texttt{hdrm} package.

\subsection{Syntax}
In this section, we outline the syntax of the two main functions provided by the package — one for the single-group setting and one for the multi-group case — in order to facilitate their application to real-world data sets.
\subsubsection{Single Group}
For analyses involving only a single group, the function \textit{hdrm\_single} has four arguments, which are:
\begin{itemize}
    \item \textit{data}: The dataset for which the test should be conducted, while they could be inserted in two different formats. One option is a single matrix, where the subjects are given as their columns. Alternatively, the data could be a vector from a data frame, which makes an additional second parameter \textit{subject} necessary.

    \item \textit{hypothesis:}
    Specifies the null hypothesis to be tested.
 For the most important hypothesis in the case of one group, the hypothesis  $\Hypo^{Flat}:\vP_d\vmu=\vnull$ can be chosen by \textit{"flat"}. More experienced users could insert an idempotent and symmetric matrix $\vT$ to investigate the null hypothesis given through $\Hypo:\vT\vmu=\vnull$. A warning is issued if the supplied matrix is not a valid projection matrix, as this may compromise the test's validity.
 \item \textit{subject}: An optional vector required only when the data is provided in vector format (e.g., from a data frame). It identifies the subject corresponding to each observation, and can consist of numeric or character values.
  \item \textit{AM}: A binary variable which specifies whether the internal calculations should be performed using an alternative hypothesis matrix. With predefined $AM=1$, this so-called companion matrix is used for the calculation of estimators and thereby reduces the required computation time. This alternative hypothesis matrix is calculated based on functions from \cite{HypoShrink}.
    
\end{itemize}

\subsubsection{Multiple Group}
The function \textit{hdrm\_grouped} for the multifactorial case has more arguments and options, namely:
\begin{itemize}
    \item \textit{data}: The data for which the test is applied, where two formats are possible for inserting the data. It could be a single data matrix, where each column represents one subject/object. 
    Moreover, the data could be entered by a vector from a data frame containing the measurements. In this case, an additional vector is required as a parameter to specify the subjects.
   \item \textit{hypothesis:} Specifies the null hypothesis to be tested.
 The most used hypotheses for the multifactorial case can be chosen through corresponding characters, which are \textit{"flat"}, \textit{"whole"}, \textit{"sub"}, \textit{"interaction"} and \textit{"identical"}.  The more experienced users could specify their hypothesis by inserting two projection matrices $\vT_W$ and $\vT_S$, building together $\vT=\vT_W\otimes \vT_S$ and testing  $\Hypo:\vT\vmu=\vnull$. If the provided matrix is not a valid projection matrix, a warning is issued, as this may affect the validity of the resulting test.

      \item\textit{group:} For both possible data formats, it is necessary to give the allocation of the measurements to the single groups, which is done with a vector containing the corresponding numbers or characters. In case of a data matrix, its length must coincide with the number of rows, while otherwise its length corresponds to the length of the data vector.

  \item \textit{subject}: An optional vector required only when the data is provided in vector format (e.g., from a data frame). It identifies the subject corresponding to each observation, and can consist of numeric or character values.

    \item \textit{AM}:  A binary variable which specifies whether the internal calculations should be performed using an alternative hypothesis matrix. With
predefined $AM=1$, this so-called companion matrix is used for the calculation of estimators and thereby reduces the required computation time. This alternative hypothesis matrix is calculated based on functions from \cite{HypoShrink}.
        
      \item \textit{subsampling:} Specifies whether subsampling-based trace estimators should be applied. If set to TRUE, all estimators use subsampling; otherwise, only estimators $B_6$ and $C_3$ do.
          
     \item \textit{B:} Number of used subsamples for the calculation of the subsampling estimators. Although it should be given as a function of $N$ (predefined is \textit{"1000*N"}), numeric values are also possible but less appropriate.

    \item \textit{seed:} An optional natural number as seed, if it should be set for reproducibility, since through the subsampling estimators, the results are slightly affected by randomness. The predefined value is no seed, and if one is used, it is removed at the end of the function call.

    \end{itemize}

\subsection{Output}
Since the differences between the outputs of the two test functions are marginal, we do not distinguish between them in the interpretation of the results.\\

The output of the functions includes several values that describe the setting, such as the dimensionality of the observations, the number of groups under comparison, and the total sample size. Additionally, it reports the specific hypothesis being tested, either by  — if a predefined option is selected — or labelled as\textit{"custom"}  in the case of user-defined hypotheses. For the specified hypothesis, the output provides the resulting value of the test statistic and the associated p-value. Since the reliability of these depends on $\hat\tau_p$ and the estimated degrees of freedom $\widehat f_P$,  both of these quantities are also included in the output. In contrast, auxiliary information such as the number of subsamples used and the number of incomplete subjects removed during preprocessing is not directly displayed in the printed output, although it is part of the returned object.

\subsection{Examples}
In this section, we illustrate the application of the two main functions provided by the \textbf{hdrm} package by analyzing a real-world data set derived from a high-dimensional repeated-measures study. To this purpose, we test different hypotheses using data from EEG-study, which was conducted at the University Clinic of Salzburg, Department of Neurology and described in detail in \cite{bathke2018}. Therein, N=160 patients of both sexes took part, all with one of four diagnoses of mental impairment: subjective cognitive complaints (SCC- and SCC+), mild cognitive impairment (MCI), and Alzheimer's disease (AD). 
For each participant, four different variables are measured (activity, complexity, mobility and brain rate) and each of them at 10 different locations of the brain (frontal left/right, parietal left/right, central left/right, temporal left/right, and occipital left/right), resulting in a dimension of $d=40$. The crossed-factor structure defined by brain location and variable allows the testing of a wide variety of hypotheses, both within and across groups. Comparisons between diagnostic categories or between sexes can yield valuable insights into neurological differences associated with these factors.

The allocation of participants across the different combinations of sex and diagnosis is summarised in Table~\ref{tab:EEG}. The resulting group sizes vary substantially, but all are considerably smaller than the dimension of the observation vector, necessitating the use of high-dimensional statistical techniques.

\begin{table}[h]
\centering
\begin{tabular}{l|c|c|c|c}
&AD&MCI&SCC-&SCC+\\
\hline
male&12&27&14&6\\
\hline
female&24&30&31&16\\
\hline
\end{tabular}
\caption{numbers of observations for different factor level combinations of sex and diagnosis.}\label{tab:EEG}
\end{table}
This data set is publicly available as part of the \textbf{HRM} package \citep{HRM}, facilitating reproducibility and further analysis by interested researchers.

\subsubsection{One Group}
Given the sample sizes, we examine whether the measurement location has an effect on any of the four variables in female subjects diagnosed with mild cognitive impairment (MCI). This corresponds to a so-called flat hypothesis, though in this case it must be explicitly specified using the hypothesis matrix $\vH=\bigoplus_{k=1}^4\vP_{10}$.

\begin{verbatim}
data(EEG)
MCIFemale=EEG[EEG$sex == "W" & EEG$group == "MCI",]
T=matrix(0,40,40)
T[1:10,1:10]=diag(1,10,10)-1/10
T[11:20,11:20]=diag(1,10,10)-1/10
T[21:30,21:30]=diag(1,10,10)-1/10
T[31:40,31:40]=diag(1,10,10)-1/10

hdrm_single(data = MCIFemale$value, hypothesis = T, subject = MCIFemale$subject)

      One Group Repeated Measure
       
Analysis of 30 subjects in 40 dimensions: 
W = 18.4195  f = 1  p.value < 1e-04 
Hypothesis type: flat \end{verbatim}
\vspace{-\baselineskip}
\verb|Convergence parameter |\textit{$\tau$ }\verb|= 1|\\

The extremely low p-value indicates strong evidence against the null hypothesis, implying that the measurement location has a significant effect on the observed values. Since the value of $\widehat\tau_P$ is equal to 1, the usage of the approximation through $K_{\widehat P}$ is applicable in this situation.

\subsection{Multi-factorial}
Next, we test for group differences among women with different diagnoses, resulting in $a = 4$ groups with unequal sample sizes. One of the primary hypotheses in multi-group repeated-measure designs is the equality of mean vectors across groups, which is available in the package as the predefined "identical" or "whole" hypothesis. The test is conducted twice, once assuming heterogeneous covariance matrices and once assuming equality of covariance matrices across groups, to compare the results.
\begin{verbatim}
data(EEG)
EEGFemale=EEG[EEG$sex == "W",]
hdrm_grouped(data = EEGFemale$value,
                       hypothesis = "whole",
                       group = EEGFemale$group,
                       subject = EEGFemale$subject,
                       subsampling = FALSE,
                       B = "1000*N",
                       seed=3141)

      Multi Group Repeated Measure
      
      
Analysis of 101 individuals in 4 groups and 40 dimensions: 
W = 0.5851  f = 3.1657  p.value = 0.2199 
Hypothesis type: whole \end{verbatim} 
\vspace{-\baselineskip}
\verb|Convergence parameter |\textit{$\tau$ }\verb|= 0.3159|\\
\begin{verbatim}
hdrm_grouped(data = EEGFemale$value,
                       hypothesis = "whole",
                       group = EEGFemale$group,
                       subject = EEGFemale$subject,
                       subsampling = FALSE,
                       B = "1000*N",
                       cov.equal=TRUE,
                       seed=3141)

      Multi Group Repeated Measure
      
Analysis of 101 individuals in 4 groups and 40 dimensions: 
W = 0.4095  f = 2.038  p.value = 0.2451 
Hypothesis type: identical \end{verbatim} 
\vspace{-\baselineskip}
\verb|Convergence parameter |\textit{$\tau$ }\verb|= 0.4907|\\

In this case, the convergence parameter $\widehat\tau_P$ is not close to 0 or 1, and therefore, possible rejections should be treated with care. But with a p-value about 0.22, for usual $\alpha$ levels, the whole-plot hypothesis cannot be rejected. Assuming the equality of covariance matrices clearly changes all values, and the p-value is even higher. This illustrates why the default setting here is heterogeneous covariance matrices.

\section{Conclusion and Outlook}

This paper introduces and explains the application of the incorporated functions in the \textsf{R} package \textbf{hdrm2}, which provides a comprehensive framework for hypothesis testing on expectation vectors in high-dimensional repeated-measure-designs. The package includes dedicated functions for both the one-group setting and multi-group scenarios, supporting a range of predefined standard hypotheses as well as user-specified general hypotheses via the corresponding hypothesis matrices.

Due to the high computational complexity typically associated with high-dimensional designs, particular emphasis was placed on efficient implementation. This is achieved through the use of subsampling estimators, which allow the number of quadratic forms to be controlled explicitly and through the exploitation of more efficient representations of hypothesis matrices. Additionally, performance-critical computations are offloaded using the \textbf{RcppArmadillo} package~(\cite{armadillo}), further reducing computation time and enhancing the applicability of the methods for more extensive datasets.

These features substantially extend the applicability of the implemented tests, while critical values derived from centred and standardised $\chi^2$ distributions provide good approximations even for small sample sizes.\\\\
To ensure continued relevance and usability, the package will undergo regular updates, incorporating new theoretical developments and computational strategies. One key direction involves extending the methodology to accommodate group-specific dimensions, complementing existing approaches and enabling numerous applications.

Moreover, while the current implementation relies on symmetric and idempotent hypothesis matrices — common assumptions that ensure uniqueness — this requirement can pose practical challenges for users unfamiliar with the underlying theory and may lead to numerical issues. Future developments will, therefore, aim to relax this constraint by adapting the quadratic forms and associated estimators to operate reliably without these assumptions.
Also, a confidence ellipsoid for $\vT\vmu$ could be incorporated, which allows further interpretations of the results. 
Lastly, the adaptation of the one-group estimators to utilise difference-based formulations — analogous to the multi-group case — offers the potential to maintain their validity under the alternative hypothesis, thereby improving the robustness and power of the test procedures.

\addcontentsline{toc}{chapter}{Bibliography}
\bibliography{Literaturnew}
\address{Paavo Sattler\\
  TU Dortmund University, Department of Statistics\\
  44221 Dortmund\\
 Germany\\}
\email{paavo.sattler@tu-dortmund.de}

\address{Nils Hichert\\
  TU Dortmund University, Department of Statistics\\
  44221 Dortmund
\\
 Germany\\}
\email{nils.hichert@tu-dortmund.de}

\end{article}

\end{document}